\documentstyle[aps,aps,epsf]{revtex}

\begin{document}

\draft
\twocolumn[\hsize\textwidth\columnwidth\hsize\csname 
@twocolumnfalse\endcsname

\title{Super-Hubble Nonlinear Perturbations During Inflation}

\author{Niayesh Afshordi$^{1,2}$ and Robert Brandenberger$^{1}$ \\ 
 \it{\small$^1$Physics Department, Brown University, Providence, RI 02912, USA }\\
 \it{$^2$\small Department of Astrophysical Sciences, Princeton University, Princeton, NJ 08544, USA \normalsize}}

\date{18 November 2000}

\maketitle

\begin{abstract}
We show that the non-linear evolution of long-wavelength perturbations  may be important in a wide class of inflationary scenarios. We develop a solution for the evolution of such nonlinear perturbations which is exact to first order in a gradient expansion. As a first application, we demonstrate that in single field models of inflation there can be no parametric amplification of super-Hubble modes during reheating. We consider the implications of the solution for recent discussions of the back-reaction effect of long wavelength perturbations on the background geometry, give a new derivation of the equation of motion of stochastic inflation, and demonstrate that if the (generalized) slow-rolling condition is not satisfied, then inevitably long wavelength vector modes for gravitational fluctuations will be generated.
\end{abstract}

\vspace*{1cm}
]   
\section{Introduction}

The hypothesis that causally unrelated regions of space-time evolve independently is one of the cornerstones in the development of relativity and one of the basic assumptions of
all relativistic theories. As an implication of this hypothesis in cosmology we expect that perturbations with long wavelengths\footnote{Throughout this paper, we use
the terms super-Hubble, infrared and long wavelength interchangeably. All refer to modes with wavelengths larger than
the Hubble radius.} evolve independently at different
points of space, as if they were independent homogeneous patches of the Universe (see e.g. \cite{Lindebook} and references therein). This expectation has been explored in the linear
theory of cosmological perturbations. In this context, it is possible to
derive integral constraints (see e.g. \cite{Traschen}) which imply that in
the case of adiabatic fluctuations there can be no generation of fluctuation modes on super-Hubble scales. However, it is possible for entropy
fluctuations to be generated on such scales, one well known example being the formation of cosmic defects during a phase transition
which leads to super-Hubble structures (see e.g. \cite{RHBrev}). Another example is the generation of axion fluctuations (see e.g. \cite{axion}). Recently, the super-Hubble range effects of entropy fluctuations have been widely studied in the context of inflationary reheating (see e.g. \cite{bassett,fabio,Lyth}). 

The aim of this work is to investigate the process of structure formation in the non-linear regime, for inflationary models based on a single scalar field. In Section 2, we show that the presence of nonlinearities is inevitable in a wide class of inflationary models.
In Section 3 we find a solution (which is exact to leading order in the gradient expansion) for nonlinear modes in the case when (but not only when) perturbations satisfy a {\it generalized slow-rolling condition}. This condition is naturally satisfied during inflation. For our solution we can define a generalized Bardeen parameter which is constant on scales larger than the Hubble radius. As a first application 
(Section 4), we demonstrate that there is no amplification of long wavelength gravitational fluctuations during reheating. In Section 5 we discuss some implications of our solution for the back-reaction of perturbations on the background geometry. In Section 6, the Langevin equation for stochastic inflation \cite{Starob} is re-derived in a way which includes the effects of gravitational (and not just matter) fluctuations,
using the results of Section 3 which can be self-consistently applied in an inflating Universe with nonlinear super-Hubble modes. 
Finally in Section 7 we demonstrate that in the case in which the generalized slow-rolling condition is violated, the generation of vector modes is inevitable.

\section{Generation of Fluctuations}

Perturbations are generated in the de-Sitter phase of an inflationary Universe as a consequence of quantum vacuum fluctuations (see e.g. \cite{MFB} for a comprehensive review). The analysis is based on the consistent quantization of the linearized metric and matter perturbations in a classical expanding background space-time. In this framework, it can be shown that the Sasaki-Mukhanov parameter $v$ \cite{Sasaki,VM1}, defined as \footnote{The cosmological scale factor is denoted by $a(t)$, the Hubble expansion rate by $H$, and the scalar matter field by $\varphi$.}:
\begin{equation}
  v \, = \, a[\delta\varphi+(\frac{\dot{\varphi}}{H})\phi] \, = \, 
  \frac{a\dot{\varphi}}{H}\zeta,
\end{equation}
in the linear regime obeys the equation of motion of a free scalar field.
Here, $\zeta$ is the Bardeen parameter \cite{Bardeen}:
\begin{equation}
  \zeta \, = \, \phi - \frac{H}{\dot{H}}(H\phi+\dot{\phi}) \, ,
\end{equation}
where $\phi$ is the variable which describes scalar metric perturbations in longitudinal gauge in a spatially flat Universe:
\begin{equation}
  ds^2 \, = \, (1+2\phi)dt^2 - a^2(t)(1-2\phi)\delta_{ij}dx^i dx^j \, ,
\end{equation}
The scalar sector is the dominant one for cosmological perturbations
generated through slow-roll inflation (see e.g. \cite{LLrev} and references therein).

The action for linearized metric and matter perturbations about a classical
expanding background becomes \cite{MFB} the action of a free scalar field $v$ with time-dependent mass (the time dependence is determined by the background geometry and is negligible on scales smaller than the Hubble radius). Hence, the perturbations can be canonically quantized, and expectation values of $v$ (and hence also of $\zeta$) can be easily calculated (see e.g. \cite{MFB}) once the state of the system is
specified. The state can be specified mode by mode in Fourier space. Our choice is the following: We fix the initial conditions for each Fourier mode when its wavelength is equal to the Planck length \footnote{The values of the variables at this time are indicated by a subscript $p$, and they are functions of $k$.}, and assume that we have vacuum expectation values at that time. This assumption is very similar to the usual choice of a vacuum state at the onset of inflation, since - at least for the usual dispersion relations \cite{BM1,MB2} - the wavefunction is almost constant on sub-Hubble scales. The advantage of our choice compared to the usual one is that we do not extrapolate the physics above the Planck energy. However, for the 
usual dispersion relations for a free field, the difference is immaterial. 
Using the equation of motion for $\zeta$ in the inflationary era,
we obtain the following expression for $<\zeta^2>$:
\begin{equation}
  <\zeta^2> \, = \,  
  \int\frac{d^3k}{(2\pi)^3}\frac{H_p^4}{2k^3\dot{\varphi}_p^2} 
  (1+\frac{k^2}{a^2H^2}) \, .
\end{equation}
Note that $\zeta$ contains the full information about the linearized metric (and matter) fluctuations (valid as long as the
slow-roll approximation holds).

Now we turn our attention to the infrared part of the spectrum
($\frac{k}{aH}\ll 1$) for which
\begin{equation}
  \phi \, \simeq \, -\frac{\dot{H}}{H^2}\zeta \, .
\end{equation}
Thus
\begin{equation}
  <\phi^2>_{IR} \, = \, \frac{\dot{H}^2}{H^4}\int_{IR}  
  \frac{d^3k}{(2\pi)^3}\frac{H_p^4}{2k^3\dot{\varphi}_p^2} \, ,
\end{equation}
where the subscript $IR$ indicates that we only include infrared modes.

Integrating the equation of motion for the background field and the Friedmann equations in the slow rolling regime gives:
\begin{equation}
  \int_{\varphi_p}^{\varphi}\frac{V d\varphi}{V^{'}} \, = \, 
  m_p^2 log(\frac{k}{a m_p}),
\end{equation}
where $m_p = (8\pi G)^{-\frac{1}{2}}$ is the Planck mass and $V(\varphi)$ is the scalar field
potential. Assuming a power law potential:
\begin{equation}
  V(\varphi) \, = \, M^4(\frac{\varphi}{m_p})^{\alpha},
\end{equation}
we get:
\begin{equation}
  \varphi_p^2 \, = \, \varphi^2 - 2m_p^2 \alpha log(\frac{k}{a m_p}) \, .
\end{equation}

Substituting in (6) and carrying out the integral gives:
\begin{equation}
  <\phi^2>_{IR} \, = \, \frac{\alpha}{48(4+\alpha)\pi^2} (\frac{M}{m_p})^4 
  (\frac{\varphi}{m_p})^\alpha
  (\frac{\varphi_i}{\varphi})^{4+\alpha},
\end{equation}
for $ |\varphi| \ll |\varphi_i|$, where $\varphi_i$ is the background field value when inflation starts.

We see that if inflation lasts long enough, i.e. if $\phi_i$ is
large enough, then $<\phi^2>_{IR}$ can become larger than one and thus infrared
perturbations can go nonlinear. As a matter of fact, although the amplitude of each mode may be small, due to the large phase space of infrared modes,
the actual value of $\phi$ in real space may be greater than 1, and so we may
not be allowed to use perturbation theory to expand the Einstein
equations. The possible importance of nonlinear effects for the evolution of space-time was pointed out in \cite{bran1,bran2}, though with a different interpretation. We will return to this issue in Section 5.

Let us now study under which conditions $<\phi^2>_{IR}$ is indeed nonlinear.
As a first step, note that the value of $M$ is constrained by the observational data.
The amplitude $A$ of a linear cosmological perturbation mode (measured in terms of $\zeta$) when it enters the Hubble radius after inflation is the same as when it exits the Hubble radius
during inflation \footnote{Although this statement is conventionally taken to be true, there are exceptions as has recently been shown for a certain class of inflationary models (see Section 4 and references quoted there).}. Consider, for example, the largest wavelength observable mode, the mode which is entering the Hubble radius today and is given by:
\begin{equation}
  k = a_{0} H_{0} \simeq
  (\frac{M}{T_{CMB}})(\frac{H_{0}}{m_p})(a_e m_p),
\end{equation}
where the subscript $0$ refers to the value of quantities today, $T_{CMB}$ denotes the present temperature of the microwave background, and
$a_e$ is the value of the scale factor when inflation ends.
Then, combining (4) and (9) gives us the amplitude of the mode:
\begin{equation}
  A = \frac{H_p^4}{\dot{\varphi_p}^2}\simeq
  (\frac{M}{m_p})^4 [log(\frac{T_{CMB} m_p}{M
  H_{0}})]^{1+\frac{\alpha}{2}}
 \, .
\end{equation}
Thus we end up with:
\begin{equation}
  \frac{M}{m_p} \simeq A^{\frac{1}{4}} [67 - \frac{1}{4} log
  A]^{-\frac{\alpha +2}{8}}.
\end{equation}
The value of $A$ is constrained by various observations, in particular by the amplitude of the cosmic microwave anisotropies on large angular scales \cite{COBE}, which give $ A \sim 10^{-10}$.

A theoretical requirement for a successful inflationary scenario is that all the
observable modes must have exited the Planck length during
inflation. The consequence is that $\varphi_{i}$ is larger than $\varphi_p$ for all the observable modes. Using (9), (12) and (13), this requirement becomes:
\begin{equation}
  \varphi_i \, > \, m_p [log(\frac{T_{CMB}
  m_p}{M H_0})]^{\frac{1}{2}} \, \simeq \, m_p [67 - \frac{1}{4} log
  A]^{\frac{1}{2}},
\end{equation}
(for this order of magnitude estimate we can set $\alpha = 1$). Inserting this constraint in (10) gives:
\begin{equation}
  <\phi^2>_{IR} \, > \, A [67 - \frac{1}{4} log
  A]^{3+\frac{\alpha}{2}} \sim 10^{-4.41 + 0.93 \alpha},
\end{equation}
at the end of inflation, for $A \sim 10^{-10}$.
We see that even if inflation starts at the latest possible
time which is allowed by theoretical/observational considerations (it is usually taken to start much earlier),
$<\phi^2>_{IR}$ almost reaches the nonlinear regime for a massless
theory (i.e. for $\alpha = 4$) just before reheating starts. If inflation starts at values of $\phi$ for which the energy density is comparable to the Planck density, then the phase space of infrared modes is much larger and the nonlinear regime will be reached much earlier.

\section{Gradient Expansion}
  
We use the following ansatz for our metric:
\begin{equation}
   ds^2 \, = \, e^{2\phi}dt^2-e^{-2\psi} \delta_{ij}dx^i dx^j,
\end{equation}
which is a generalization of a metric with linear scalar perturbations. This ansatz, though not a general solution, represents the most important sector of the metric since any such metric can be continuously connected to a metric with linear scalar perturbations which is the dominant sector in a Universe with linear perturbations generated during slow-roll inflation \cite{LLrev}.
We also consider the simplest model of matter which has one scalar field (the inflaton).

The Lagrangian and energy-momentum tensor are given by:
\begin{eqnarray}
 {\mathcal{L}} = \frac{1}{2} \partial^{\alpha}\varphi \partial_{\alpha}\varphi -
  V(\varphi), \\
  T^{\mu}_{\nu} = \partial^{\mu}\varphi\partial_{\nu}\varphi -
  \mathcal{L}\delta^{\mu}_{\nu}.
\end{eqnarray}

In writing the Einstein equations we keep the leading terms in their gradient expansions, assuming
that the wavelengths are large. Naturally, in the $G_{00}$ and $G_{ij}$ equations, all the gradient terms
drop out \footnote{It is possible to show that the $G_{ij}$ equations for $i\not= j$ can be satisfied only
if we include the vector and tensor sectors of the metric perturbations. However their magnitudes go to zero in
the long wavelength limit and hence, in this limit, they do not affect the equations for scalar perturbations.}  
and we are left with the Friedmann equations for the above metric:
\begin{eqnarray}
   {\psi^{\prime}}^2 = \frac{ 8\pi
   G}{3}(\frac{{\varphi^{\prime}}^2}{2}+V(\varphi)), \\
   -2\psi^{\prime\prime}+3{\psi^{\prime}}^2 = 8\pi G(
   -\frac{{\varphi^{\prime}}^2}{2} + V(\varphi)),
\end{eqnarray}
and of course the more interesting $G_{0i}$ equation:
\begin{equation}
   (\psi^{\prime})_{,i} = 4\pi G \varphi^{\prime} \varphi_{,i}
\end{equation}
where $^{\prime}$ denotes $e^{-\phi}
\frac{\large\partial}{\partial t \normalsize}$.

We immediately see that not every $\varphi$ can satisfy the last equation since the right hand side must
be a perfect gradient. The condition that $\varphi$ must satisfy is:
\begin{equation}
  (\varphi^{\prime})_{,i}\varphi_{,j} \, = \, 
  (\varphi^{\prime})_{,j}\varphi_{,i} \, ,
\end{equation}
or, equivalently\footnote{Equation (22) implies that the cross product of two gradient vectors vanishes, so
the vectors must be in the same direction.}:
\begin{equation}
  \mathbf{\nabla}\varphi^{\prime} \, \propto \, \mathbf{\nabla} \varphi
\end{equation}

From Equation (23), we can see that surfaces of constant $\varphi^{\prime}$ and constant $\varphi$
must be tangential to each other and consequently are the same. This means that $\varphi^{\prime}$ must be a function
of $\varphi$, at least in the finite regions of space in which the gradient of $\varphi$ does not vanish:
\begin{equation}
   \varphi^{\prime} \, = \, \frac{\partial}{\partial \varphi}g(\varphi,t),
\end{equation}
where $g(\varphi,t)$ can be any function of $\varphi$ and $t$. Then the $G_{0i}$ equation reduces to \footnote{
The integration constant can be absorbed in $g(\varphi,t)$.}:
\begin{equation}
    \psi^{\prime} \, = \, 4\pi G g(\varphi,t).
\end{equation}
Since there is no explicit time dependence in this formalism, we assume that $g$ has also no explicit time dependence:
\begin{equation} \label{srcond}
   g(\varphi,t) \, = \, g(\varphi).
\end{equation}
From now on, we call Equation (24) (with no explicit time dependence) the generalized slow-rolling condition due to its resemblance to the field equation during the slow-rolling phase. Note that at this point, we have not shown that the absence of time dependence is required in (\ref{srcond}), but simply that this ansatz gives a consistent solution. However, below we show that in the case of single field inflationary models which initially satisfy the slow rolling condition, (\ref{srcond}) is indeed satisfied.

As one can easily see, in the linear regime Equations (25) and
(26) lead to the vanishing of the non-adiabatic pressure defined as:
\begin{equation}
  \delta p_{nad} \, \equiv \, \dot{p} \Gamma \equiv \dot{p}
  (\frac{\delta p}{\dot{p}} -\frac{\delta \rho}{\dot{\rho}}) \, .
\end{equation}
This is the condition derived in \cite{fabio,Lyth} for the constancy of $\zeta$ for the perturbations. 

Combining Equations (24) and (26) gives:
\begin{equation}
   \psi^{\prime\prime} \, = \, 4 \pi G {\varphi^{\prime}}^2,
\end{equation}
which can also be obtained by combining the Friedmann equations (19) and (20).

After neglecting the gradient terms and using Equations (24) and (26),
both the $G_{00}$ and $G_{ij}$ equations reduce to:
\begin{equation}
   4\pi G \ {g^{\prime}}^2 - 3(4\pi G \ g)^2 \, = \, -8\pi G \ V(\varphi).
\end{equation}

Thus we see that, in the long wavelength limit, the $G_{0i}$ equation is consistent with the Friedmann equations 
given that the field $\varphi$ satisfies the generalized slow-rolling condition (24).

As a matter of fact, this is what happens during slow-roll inflation, 
when we have:
\begin{equation}
   \varphi^{\prime} \, = \, - \frac{1}{3H}\frac{\partial 
   V(\varphi)}{\partial  \varphi} \, = \, 
   \frac{1}{3\psi^{\prime}}\frac{\partial V(\varphi)}{\partial \varphi}  \, ,
\end{equation}
and hence (24) is satisfied.

\begin{figure}
\epsfxsize=2.9 in \epsfbox{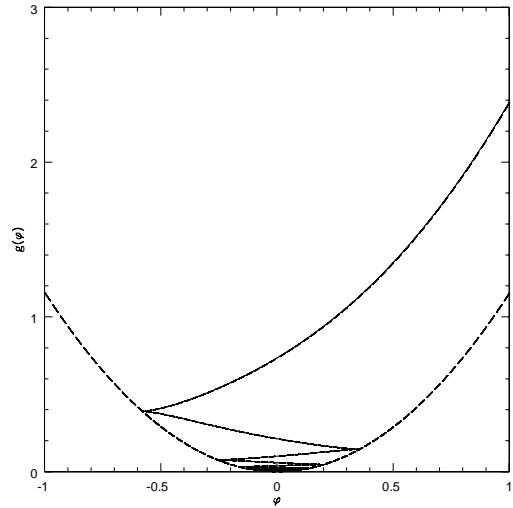}
\caption{The solid curve is a solution for $g(\varphi)$ during the reheating
era and the dashed curve is $\sqrt{V(\varphi)/(6 \pi G) }$. The horizontal
axis is in units of Planck mass while the vertical axis has arbitrary
units.}
\label{fig1}
\end{figure}

Because of this property, even after the end of inflation, as long as the perturbation
is in the super-Hubble regime, $\varphi^{\prime}$ remains a function of $\varphi$. The reason is that, using the field equation,
we can track the evolution of $\varphi$  and $\varphi^{\prime}$ into the reheating era, as functions of the
initial $\varphi$ during inflation. By eliminating the initial $\varphi$, it is possible to find $\varphi^{\prime}$  as a
function of $\varphi$ at any time. However it will not be a single-valued function when $\varphi$ is oscillating. Figure (\ref{fig1}) shows the behavior of  $g(\varphi)$ (solid curve) during the reheating era,
obtained by solving (29). The dashed curve is $\sqrt{V(\varphi)/(6 \pi
G) }$, the asymptotic limit of $g(\varphi)$ for $\varphi \gg m_p$.

Dividing Equation (24) by Equation (25) and integrating the result with respect to $\varphi$, we find that:
\begin{equation}
  \tilde{\zeta} (x) \, \equiv  \, \psi - 4\pi G 
  \int (\frac{\partial  \log(g)}{\partial \varphi})^{-1} d\varphi,
\end{equation}
is an integration constant and as such does not depend on time.

It can easily be verified that for linear perturbations 
$\delta \tilde{\zeta}(x)$ reduces to the usual Bardeen parameter $\zeta$, which is known to remain constant
for adiabatic super-Hubble modes \footnote{In linear perturbation theory, $\delta\phi$ is forced to be equal to $\delta\psi$ by the off-diagonal $G_{ij}$
equations for all theories in which the off-diagonal components of $T_{ij}$ vanish to linear order, which is in particular the case for scalar field matter.}:
\begin{eqnarray}
 \delta \tilde{\zeta}(x) \, & = & \, \delta \psi - 
 4\pi G \frac{g \delta \varphi}{g^{\prime}} \, = \, \delta \psi - 
 \frac{g\delta{(e^{-\phi}\dot{\psi})}}{{g^{\prime}}^2} \nonumber \\
 & = & \, \delta \psi - 
 \frac{H}{\dot{H}}(H\delta\phi+\delta\dot{\psi})=\zeta(x).
\end{eqnarray}
Thus, we have been able to generalize the Bardeen parameter to the case of nonlinear long-wavelength fluctuations.
 
Let us now consider the case of multiple scalar fields. In this case,
Equation (21) changes to:
\begin{equation}
 (\psi^{\prime})_{,i} = 4\pi G \varphi_{a}^{\prime} \varphi_{a,i},
\end{equation}
where the subscript $a$ denotes the field index and is summed over,
if repeated. The consistency condition (22) takes the form:
\begin{equation}
 (\varphi_a^{\prime})_{,i}\varphi_{a,j} \, = \, 
 (\varphi_a^{\prime})_{,j}\varphi_{a,i}.
\end{equation}
We see that if the fields satisfy the generalized slow rolling condition
\begin{equation}
  \varphi_a^{\prime} \, = \, \frac{\partial}{\partial \varphi_a}g(\varphi,t),
\end{equation}
then (34) is satisfied and we again arrive at (25). However, for the 
multiple field problem this result is not general. Note that (35) is
only a sufficient condition. 

During inflation, if the effective mass of a field component is 
larger than the Hubble constant, it is damped in a few e-foldings.
Otherwise, it slow-rolls down the potential. So a few e-foldings
after the onset of inflation, all the remaining field components
undergo slow-rolling and satisfy (35). We conclude that during
multiple-field inflation, we can still use the above results. 
However when the inflation ends, (35) breaks down and, unlike 
the single field case, we end up with a mixture of scalar
and vector metric perturbations, as described in Section 7.  
 
We are also able to define the generalized Bardeen parameter (31)
for each slow-rolling path in field space, since the evolution
is effectively of single field nature along each path. However, the parameter
does not remain constant after inflation ends since (35) breaks down.
 
In conclusion, we have shown that causally disconnected regions of the
Universe evolve independently in the sense that locally observable
parameters satisfy the conservation law (31),
which is a generalization of the conservation of the Bardeen
parameter in linear perturbation theory. As mentioned, our result is only true in the case of (in general) non-linear scalar perturbations satisfying the slow-roll condition.

In the next two sections we consider some of the applications of
this result.

\section{Parametric Resonance of Super-Hubble Modes during Reheating}

It was recently suggested \cite{Kaiser} that parametric resonance
during the reheating phase of an inflationary Universe \cite{TB} may lead to
an exponential amplification of super-Hubble scale gravitational
perturbations. If true, this would affect the usual predictions of
inflationary model for observables such as the matter power
spectrum and the spectrum of cosmic microwave anisotropies.

In Ref. \cite{finelli} it was shown that, although there is no
causality constraint which prohibits the amplification of
super-Hubble ( but sub-horizon) modes during reheating, the effect
does not occur in a simple massive scalar field model of chaotic
inflation (i.e. with $\alpha = 2$ ). This result was shown to be true even
beyond the linear analysis, using numerical methods \cite{parry}. Recently, a
general no-go theorem for resonance of long wavelength scalar gravitational 
fluctuations in the context of a single scalar matter field theory
was suggested in \cite{lin}. In this last reference, the effect was also investigated numerically for a matter theory with both quadratic and quartic
terms (see also \cite{easther}).

The problem with the analysis of \cite{lin} or with every other analytical approach which is based on considering the evolution of $\zeta$ (in the linear regime) through the turning point of $\varphi$ is that $\zeta$ is ill-defined
when $\dot{\varphi}$ vanishes.
However, $\tilde{\zeta}$ is continuous at this point since the integrand of (30) diverges as
$(\varphi - \varphi_0)^{-\frac{1}{2}},$\footnote{This can be obtained by considering Equation (21) close to the turning point.}
so the integral is continuous. Therefore, $\tilde{\zeta}$ is well defined through the turning. We now argue that $\tilde{\zeta}$ is in fact constant throughout.
For the argument we refer back to Figure 1 which shows a sketch of the time evolution of $g(\varphi)$ during the period of oscillation of $\varphi$. During each time interval between the turning points, $\tilde{\zeta}$ and hence $\zeta$
are constant. At the turning points, $\tilde{\zeta}$ is continuous, and hence
it is not possible that at these points $\tilde{\zeta}$ undergoes a jump.
The fact that $\tilde{\zeta}$ is constant implies the absence of
parametric amplification of super-Hubble modes.

In contrast to the single field case, as pointed out at the end of Section 3, this formalism cannot be applied to the multiple-field case during reheating, and thus we 
may expect non-trivial behavior of the fields during this era. In fact, such behavior has been observed for a 
specific two-field potential \cite{bassett,fabio,zibin,gordon}, where parametric amplification of isocurvature
modes can lead to non-linear amplitudes of infrared modes. We will return to this effect in Section 7.

\section{Back-Reaction of Adiabatic Infrared Modes}

It has been claimed \cite{bran1,bran2} that the growth of $<\phi^2>$ during slow-roll inflation can lead to significant corrections to the background
Friedmann equations at second order in perturbation theory (see also \cite{RW1,RW2} for similar discussions based on the back-reaction of infrared gravitational waves). We claim that, as a result of the analysis of
Section 3, this effect, though formally valid, cannot be identified
by local observers
as an effect of inhomogeneities. The reason is
that, as we showed, for perturbations generated
during slow-roll inflation, the local Friedmann equations are
always satisfied. Note that our analysis includes the effects of the leading back-reaction terms in the abovementioned references. Obviously, back-reaction terms which are of higher order in gradients are not included in our analysis.

Due to the fact that inflation is followed by a radiation
dominated era, modes which enter the Hubble radius after 
inflation are damped by the Hubble expansion. Consequently, the
phase space of constant amplitude perturbations is  
shrinking after inflation. This effect can smooth
out the density field so that it becomes of linear order again at 
some point in the radiation dominated era. This will happen in
models in which inflation does not last much longer than the minimal
number of e-foldings required to solve the problems of standard big bang
cosmology. However, in many models of chaotic inflation, inflation lasts long
enough, so that the remaining phase space at the time of equal matter and radiation is large enough to give nonlinearity. However, even in this
case there has not been enough time for the non-linearities to
enter the observable region (sub-Hubble scales). 

This does not
mean, however, that the presence of nonlinearities has no effect at all.
As we will see in the next section, the process of generation
of perturbations during inflation is influenced by nonlinear
effects and thus the spectrum is distorted compared to what would be
predicted in linear theory.

\section{Equation of Motion for Stochastic Inflation}

In this section we will use coarse graining on the scale of the Hubble radius to provide a new derivation of the equation of motion for stochastic inflation \cite{Starob} which takes into account the gravitational fluctuations. 

Let us define the coarse-grained Bardeen parameter, $\zeta_c$, in Fourier space as follows:
\begin{equation}
  \zeta_c(k) \, = \, \tilde{\zeta}(k) W(\frac{k}{aH})
\end{equation}
where the window function $W(k)$ can be chosen to be\footnote{This is only
one example of a cutoff function. In general, one can use any function which is 
close to one for $k < aH$ and tends to zero as $k$ goes to infinity fast enough to eliminate the ultraviolet divergence.}:
\begin{equation} 
 W(\frac{k}{aH}) \, = \, (1+\frac{k^2}{a^2 H^2})^{-2} 
\end{equation}
which is smooth on the Hubble scale. In a patch larger than the Hubble radius but small
enough so the non-linearity in $\zeta$ is negligible, we can use $\zeta$ 
(usual Bardeen parameter) instead of $\delta \tilde{\zeta}$. Making use of the fact that background $\tilde{\zeta}$ is constant in this patch 
we have:
\begin{equation}
 \dot{\tilde{\zeta}} \, = \, \delta \dot{\tilde{\zeta}} = \dot{\zeta}.
\end{equation} 

We can use our knowledge of the linear quantum generation of perturbations
(Section 2) to find the statistical properties of  $\tilde{\zeta}^{\prime}$,
with the result
\begin{eqnarray}
&\,& <\tilde{\zeta_c}^{\prime}({\mathbf{x_1}},t_1) \tilde{\zeta_c}^{\prime}({\mathbf{x_2}},t_2)> \\
 &=& \, \frac{5 H^6}{48 \pi^2 \varphi^{\prime^2}} 
F(H a |{\mathbf{x_1-x_2}}|,H|t_1 - t_2|) \, , \nonumber
\end{eqnarray} 
where $F(0,0) = 1$ and $F(\beta,\tau) \rightarrow 0$ as $ \beta $ or $ \tau $ 
goes to infinity (more precisely, $F(\beta, \tau) \ll 1$ if $\beta \gg 1$ or
$\tau \gg 1$). The explicit form of $F$ is complicated and we do not write it down here. 

Taking the time derivative of the definition of $\tilde{\zeta}$ (see equation (31)), and using the fact that $\psi' = - H \simeq - \sqrt{{{8 \pi G} \over 3} V(\varphi)}$ during inflation, we get
\begin{equation}
\tilde{\zeta}' \, = \, - H - 4 \pi G \bigl({{\partial {\rm log} (g)} \over {\partial \varphi}} \bigr)^{-1} \varphi' \, . \nonumber
\end{equation}
We can neglect $g'$ in (29) to find $g$ as a function of $V(\varphi)$ in the slow-roll approximation. Plugging this into the above formula, we obtain the coarse-grained equation of motion:
\begin{equation}
 \varphi^{\prime} \, = \, -m^2_p H \frac{\partial \log(V)}{\partial \varphi} - \sqrt{\frac{5}{48}}\frac{H^2}{\pi}\xi(x,t) \, .
\end{equation}  
Here, all quantities are coarse-grained, and the variable $\xi$ is a random Gaussian field whose two-point correlation function is given by:
\begin{equation}
<\xi({\mathbf{x_1}},t_1) \xi({\mathbf{x_2}},t_2)> 
\, = \, F(H a |{\mathbf{x_1-x_2}}|,H|t_1 - t_2|) \, .  
 \end{equation} 
Note that since $\zeta$ is a metric perturbation variable, the above equations correspond to metric back-reaction yielding stochastic dynamics.

Let us compare this calculation with the original one in \cite{Starob}. There, instead of the window function (37), a step function was used. More importantly, gravitational perturbations were neglected.      
As a result, the numerical factors are different. In addition, in the original analysis the random variable had a white noise spectrum whereas
our noise has a finite correlation time. In spite of these differences, we expect that the qualitative behavior of the solutions 
will stay the same. 

One major difference between our analysis and the usual analysis of stochastic inflation (see e.g. \cite{Linde2})
is that if inflation stops at some point, and consequently the generalized slow rolling condition breaks down (see the end
of Section 3), then, as elaborated in Section 7, the evolution takes a nonlocal form. In particular, the end of inflation
in some region of space can affect the inflating regions in an acausal way. 

Limiting the analysis to the linear regime leads to the standard results of the stochastic inflationary scenario. Namely, since
in the large picture different points in space undergo independent random evolution, we end up with a scale invariant
spectrum for $\zeta$ (or $\tilde{\zeta}$) with a logarithmic correction in Fourier space due to the finite 
correlation length of the random field $\xi$. 

However, as argued in Section 2, there is a large class of scenarios which
allow infrared nonlinear perturbations. As a matter of fact, the main result of Section 2 is that the cumulative effect
of the second term in (41) may become important even when its magnitude is negligible. Hence, even when the self-reproduction of the stochastic scenario does not take place, infrared nonlinearities may be important \cite{bran1}.  

This nonlinearity
affects the generation of perturbations by changing the background parameters which appear in the amplitudes as well as
the Gaussianity of the inhomogeneities. We postpone the calculation of this correction to a future work.   
     
\section{Vector Modes}

In this section we investigate the case in which the generalized slow-rolling condition (24) or (35) is
not satisfied. For the general metric:
\begin{equation} 
 ds^2 \, = \, N^2 dt^2 - \gamma_{ij} dx^i dx^j, 
\end{equation}
the $0i$ Einstein tensor elements are given by
\begin{equation} \label{zeroi}
 G^0_i \, = \, N^{-1}(K_{,i} -K^j_{i;j}),
\end{equation} 
where 
\begin{eqnarray} 
 K^i_j \, &=& \, -\frac{1}{2}N^{-1}\gamma^{ik} \dot{\gamma}_{kj}, \\
 K \, &=& \, K^i_i.
\end{eqnarray} 
For the metric used in Section 3:
\begin{equation}
 N = e^{\phi}, \gamma_{ij} = e^{-2\psi}\delta_{ij}
\end{equation}
and thus:
\begin{equation}
 K_{ij} \, = \, \psi^{\prime} \gamma_{ij} \, .
\end{equation}

The $G^0_i$ equation is not satisfied in general. It is well-known that beyond linear order in perturbation theory, tensor and scalar perturbations mix. However, in our case we cannot satisfy the equation by adding tensor perturbations. We thus include vector metric perturbations in order to satisfy the $G^0_i$ equation.

The presence of vector modes would modify $K$ as follows:
\begin{equation}
 K_{ij} \, = \, \psi^{\prime}\gamma_{ij} + A_{i;j}+A_{j;i},
\end{equation}
where $A_i$ is some vector field made up from the metric variables representing vector perturbations. Substituting this into (\ref{zeroi}) leads to the 
following $G^0_i$ equation:
\begin{equation} \label{comp}
  2{\mathbf{\nabla}}\psi^{\prime} - 2{\mathbf{R}}.{\mathbf{A}} +
  {\mathbf{\nabla\times(\nabla\times A)}} \, = \, 8 \pi G \varphi^{\prime}
  {\mathbf{\nabla}} \varphi
\end{equation}  
where $\mathbf{R}$ is the Ricci tensor. If $\mathbf{R}$ vanishes, it is easy to find $\mathbf{A}$ and $\psi^{\prime}$
in Fourier space:
\begin{eqnarray} \label{endeq}
  {\mathbf{A_k}} \, &=& \, -\frac{8 \pi G}{k^4} {\mathbf{ k\times(k\times C_k)}},\\
  \psi_k^{\prime} \, &=& \, \frac{-4i\pi G}{k^2} ({\mathbf{k.C_k}}),
\end{eqnarray}   
where
\begin{equation}
  {\mathbf{C_k}} \, \equiv \, 8 \pi G\int d^3x e^{-i{\mathbf{k.x}}} \varphi^{\prime}{\mathbf{\nabla}} \varphi,
\end{equation}
is the Fourier transform of the right hand side of (\ref{comp}).

One can take the $\mathbf{R}$ term to the right hand side of (\ref{comp}) and write a perturbative expansion in $\mathbf{R}$ for $\mathbf{A}$ and $\psi^{\prime}$. We conclude that for general nonlinear scalar field perturbations the metric which satisfies
the Einstein equations, in general, must include nonlinear vector perturbations.  
One known example of this phenomenon is the formation of vector perturbations in the case of topological defect formation during a phase transition, where super-Hubble structures form in a short time \cite{RHBrev}. Another example considered in \cite{bassett,fabio,zibin} (see the end of Section 4) occurs in inflationary reheating, since during
multiple field reheating the generalized slow-rolling condition (35) breaks down. Note that in the back-reaction calculation of \cite{zibin} and in models in which the infrared modes lead to nonlinearities, it is not enough to include only scalar perturbations since vector perturbations are generated as well.   
 
\section{Conclusion}

We have demonstrated that in many models of inflation, the large phase space
of infrared modes leads to nonlinearities, even when the amplitude of each
Fourier mode is small. We were able to find a solution of the Einstein field equations for nonlinear fluctuations which is exact to leading order in the gradient expansion, and thus will be accurate to describe the infrared modes. For this solution, we were able to define a generalized Bardeen parameter which is conserved in time.

As a first application of this formalism, we were able to show that in models with a single matter field, there can be no parametric amplification of super-Hubble cosmological fluctuations during inflationary reheating. Applied to the problem of back-reaction of infrared modes, our solution implies that the back-reaction effect is not locally identifiable as an effect due to inhomogeneities since the local Friedmann equations are satisfied. We were able to use our formalism to give a re-derivation of the equation of motion for stochastic inflation which takes into account the effects of gravitational fluctuations. Finally, we have shown that nonlinearities inevitably lead to
the generation of vector perturbations if a {\it generalized slow-rolling condition} is not satisfied.
   
This work has been supported in part by the US Department of Energy under Contract DE-FG02-91ER40688, Task A.

\end{document}